\documentclass[prl,showpacs]{revtex4}
\begin{document}

\newcommand\beq{\begin{equation}}
\newcommand\eeq{\end{equation}}
\newcommand\bea{\begin{eqnarray}}
\newcommand\eea{\end{eqnarray}}
\title{Number Fluctuation and the Fundamental Theorem of 
Arithmetic}
\author{ Muoi N. Tran and Rajat K. Bhaduri } 
\affiliation { Department of Physics and Astronomy, McMaster University,
Hamilton, Ontario, Canada L8S 4M1}

\begin{abstract}
We consider $N$ bosons occupying a discrete set 
of single-particle quantum states in an isolated trap. 
Usually, for a given excitation energy, there are many combinations of 
exciting different number of particles from the ground state, resulting in 
a fluctuation of the ground state population. As a counter  
example, we take the quantum spectrum to be logarithms of the prime number 
sequence, and using the fundamental theorem of arithmetic,  
find that the ground state fluctuation vanishes exactly for 
{\it all} excitations. The use of the standard canonical or 
grand canonical ensembles, on the other hand, gives substantial number 
fluctuation for the ground state. This difference between the microcanonical 
and canonical results cannot be accounted for within the framework of 
equilibrium statistical mechanics.

\end{abstract}
\vskip .5 true cm

\maketitle
%\narrowtext

After the experimental discovery of BEC in a trapped dilute gas at ultra-low 
temperatures, much attention has been paid to the problem of number 
fluctuation in the ground state of the ideal system~
\cite{grossman1,politzer,grossman2,weiss,gazda,herzog,navez,
holthouse,holthaus}, as well as a weakly interacting bose 
gas~\cite{buffet,stringari,illuminati,idziaszek,xiong1,xiong2,xiong3,bhaduri}.
There are also a few papers on ground state fluctuations in a trapped fermi 
gas~\cite{tran}. There are several reasons for this interest. In standard 
statistical mechanics, number fluctuation is related to density-density 
correlation, and to the compressibility of the system in the 
grand canonical ensemble (GCE)~\cite{bhaduri}.
Light scattering cross section off the medium, in principle, may be 
related to the ground state fluctuation in the system~\cite{light}. 
The so called grand canonical catastrophe in an ideal bose 
gas, where the fluctuation diverges at low temperatures, was already 
known~\cite{fierz}. Therefore a more 
accurate treatment of the problem was needed for trapped gases. 
In the microcanonical 
treatment of number fluctuation from the ground state in a harmonic trap, the 
problem is closely related to the combinatorics of partitioning an 
integer, and thus there was an interesting link to 
number theory~\cite{grossman1}. It turned  
out that the result for the ground state number fluctuation was very 
sensitive to 
the  kind of asymptotic approximations that are made. Another aspect that 
drew much attention in the literature was the difference in the calculated 
results for fluctuation using the canonical and the microcanonical 
formulations~\cite{navez,holthouse}. It was pointed out by 
Navez {\it et al.}~\cite{navez}  
that for a trapped bose gas below the critical temperature, the 
microcanonical result for fluctuation could be obtained solely using the 
canonically calculated quantities, which in turn may be obtained from the 
so called Maxwell Demon ensemble~\cite{martin}, to be explained later.       

In this paper, we give an example of a quantum spectrum that has 
no number fluctuation in the ground state for any excitation energy in 
the microcanonical ensemble, as a 
direct corollary of the fundamental theorem of arithmetic. 
The canonical ensemble, on the other hand, yields a dramatically different 
ground state number fluctuation. The method of Navez {\it et al.}  
fails to account for the difference between the microcanonical and canonical 
results in this example. 
This failure of the canonical (grand canonical) ensembles 
is due to the peculiar nature of the single-particle spectrum in our example. Generally, when a large excitation 
energy is supplied to a system, there are a very large number of distinct 
microscopic configurations that are accessible to it. 
All these different microstates describe the same macro-state of a given 
excitation energy. The classic 
example is that of bosons in a harmonic trap, where the number of partitions 
of an integer number, corresponding to the number of microstates, increases 
exponentially. We use, on the other hand, 
another example from number theory, to propose a system where {\it the 
excitation energy, no matter how large, is locked in one microstate}. 
Consequently, although it is possible to explicitly calculate the canonical 
or grand canonical partition functions and therefore the  thermodynamic 
entropy for this example, it does not approach or equal the information 
theoretical entropy that can be exactly calculated using number 
theory~\cite{planes}. So far as we know, this constitutes a novel and new 
example that links number theory and thermodynamics, and gives radically 
different results for the microcanonical and canonical ensembles.

We consider bosons in a hypothetical trap with a single-particle 
spectrum ({\it not} including the ground state, which is at zero energy)
\beq
\epsilon_p=\ln p~,
\label{primes}
\eeq
where $p$ runs over the prime numbers $2,3,5,..$. Of course, such a spectrum 
is not realizable experimentally, and it is merely a means of 
performing a thought experiment. We shall use, in what follows, both a 
truncated sequence of primes, as well as the infinite sequence when we perform 
the canonical calculations for fluctuation. First, however, we perform 
the exact calculation for number fluctuation from the ground state. 
Suppose that there are $N$ bosons 
in the ground state at zero energy, and an excitation energy $E_x$ is given 
to the system. In how many ways can this energy be shared amongst the bosons 
by this spectrum ? Before giving the answer, we remind the reader of the 
fundamental theorem of arithmetic, which states that every positive integer 
$n$ can be written in only one way as a product of prime numbers~
\cite{edwards} :  
\beq
n=p_1^{n_1}p_2^{n_2}...p_r^{n_r}...~,
\label{funda}
\eeq 
where $p_r$'s are distinct prime numbers, and $n_r$'s are positive integers 
including zero, and need not be distinct. It immediately follows from 
Eq.~(\ref{funda}) that if the excitation energy $E_x=\ln n$, where the 
integer $n\geq 2$, 
there is only one unique way of exciting the particles from the ground state. 
If $E_x\neq \ln n$, the energy is not absorbed by the quantum system. Since 
the number of bosons excited from the ground state, for a given $E_x$, is 
unique for this system, {\it the number fluctuation in the ground state is 
identically zero} ! Moreover, this conclusion is valid whether we take in    
Eq.~(\ref{primes}) an upper cut-off in the prime number, or the infinite 
sequence of all the primes. The information-theoretic entropy at an 
excitation energy $E_x$ is 
\beq
{\cal{S}}(E_x)=-\sum_i P_i \ln P_i~,
\eeq
where $P_i$ is the probability of excitation of the microstate $i$. Since 
only one microstate contributes with unit probability, and all others have 
$P_i=0$ (when $ E_x=\ln n$), the entropy ${\cal{S}}=0$. It is also straight
forward to calculate the ground state population $N_0$ as a function of the 
excitation energy $E_x$. For this purpose, we truncate the spectrum given 
by Eq.~(\ref{primes}) to the first million primes, with a cutoff denoted by 
$p^{\star}$, and take $N=100$.  
We shall display the numerical results after describing the canonical 
calculations.

Our next task is to calculate these quantities in the canonical ensemble, 
and see if the differences in the microcanonical and canonical results 
may be accounted for (as $N\rightarrow \infty$) using the method of 
Navez {\it et al.}~\cite{navez}. We first do the calculation for the 
truncated spectrum. The one-body canonical partition function is then given 
by $Z_1(\beta)=1+\sum_{p=2}^{p^{*}}
 \exp (-\beta \ln p)$. The N-body bosonic  
canonical partition function is obtained by using the recursion 
relation~\cite{borrmann} 
\beq
Z_N(\beta)={1\over N}\sum_{s=1}^N Z_1(s\beta) Z_{N-s}(\beta)~.
\eeq
Once $Z_N$ is found, the ground state occupation $N_0=\left<n_0\right>_N$ and the 
ground state number fluctuation for the canonical 
ensemble can be readily be computed ~\cite{tran,parvan}. We define 
$\delta N_0^2=(\left<n_0^2\right>_N-\left<n_0\right>_N^2)$, where the RHS is calculated using 
Eqs.~(24) and (25) of Ref.(\cite{tran}). In Fig.~1a) and b), we display 
the results of the canonical calculations for the ground state occupancy 
fraction $N_0/N$ and the ground state fluctuation 
$(\delta N_0^2)^{1/2}/N$ for $N=100$ as a function of temperature $T$ with 
the truncated spectrum of the first million primes. 
%\begin{figure}[!t]
%\includegraphics[width=2.6in,angle=270]{/home/tran/PRIME/Fig1a.ps} 
%\includegraphics[width=2.6in,angle=270]{/home/tran/PRIME/Fig1b.ps} 
%\caption{a) Average occupancy in the ground state $N_0/N$ versus temperature $T$ for $N=100$ in the canonical and grand canonical ensembles. b) Plot of the relative ground state number fluctuation in both ensembles.  Note the steep rise in the grand canonical fluctuation.}
%\end{figure}
For comparison, we also show 
the results of the corresponding grand canonical calculation. The grand 
canonical catastrophe for the number fluctuation is clearly evident. 
It is also easy to calculate 
the canonical (equilibrium) entropy $S=\ln Z_N(\beta)+\beta E_x$.
The comparison with the combinatorial (or microcanonical) results requires 
that we convert the canonical temperature to excitation energy $E_x$. This is 
done by using the standard relation $E_x=-{\partial \ln Z_N(\beta)\over 
{\partial \beta}}$. In Fig.~2, the canonical fractional occupancy of the 
bosons in the excited states, $\left<N_e\right>/N$, for $N=100$, is compared with the 
(exact) combinatorial (or microcanonical) calculation as a function of the 
excitation energy $E_x$. Although the 
canonical and the grandcanonical $\left<N_e\right>/N$ are nearly identical, the 
corresponding microcanonical quantity is radically different. This anomaly 
persists even as $N\rightarrow\infty$, showing the breakdown of the 
equivalence between the microcanonical and the other ensembles.  
In Fig.~3, we display the behaviour of the canonical entropy as a 
function of $T$ and $E_x$.
The microcanonical entropy ${\cal S}(E_x)$, of course, is zero, and so is 
the number fluctuation $\delta N_0^2$. Thus we find that the canonical results 
have no resemblance with the exact microcanonical ones. All these calculations 
were performed for a truncated spectrum (first million primes) of $\ln p$, as 
specified earlier, and for $N=100$. 

%\begin{figure}[!t]
%\includegraphics[width=2.6in,angle=270]{/home/tran/PRIME/Fig2.ps}
%\caption{a) Plot of the average occupancy in the excited states $N_{ex}/N$, for $N=100$, versus the excitation energy $E_x$ in the canonical ensemble (continuous bold curve), compared with the exact microcanonical calculation. The latter calculation is done for $E_x=\ln n$, where $n$ is an integer, and the results are shown by dark points. These are joined by dotted lines to emphasize their zigzag character. For example, the sixth point (including $0$) corresponds to $E_x=\ln 6$, and gives $N_e=2$, corresponding to the prime factor decomposition $2\times 3$. }
%\end{figure}

We shall now use the procedure of Navez {\it et al.}~\cite{navez} to check 
if the microcanonical results may be obtained from a canonical calculation 
as $N\rightarrow \infty$. These authors constructed the so called Maxwell's 
Demon ensemble in which the ground state (for $T<T_c$) was taken to be the 
reservoir of bosons 
that could exchange particles with the rest of the subsystem (of the excited 
spectrum) without exchanging energy. Denoting the grand canonical partition 
function of the excited subsystem by $\Xi_e(\alpha, \beta)$, with 
$\alpha=\beta\mu$, it was shown that the canonical occupancy of the excited 
states, $\left<N_e\right>$, and the number fluctuation $(\delta^2 N_e)$ could be 
obtained from the first and the second derivative of $\Xi_e$ with respect to 
$\alpha$, and then putting $\alpha=0$. 
%\begin{figure}[!t]
%\includegraphics[width=2.6in,angle=270]{/home/tran/PRIME/Fig3.ps}
%\caption{Plot of the canonical entropy as a function of temperature $T$ on the left, and excitation energy $E_x$ on the right, for $N=100$. The microcanonical entropy is zero. }
%\end{figure}
It was further noted that the microcanonical number 
fluctuation for the excited particles was related to the canonical 
quantities by the relation 
\beq
\langle\delta^2 N_e\rangle_{MC}^{\infty}=
\langle\delta^2 N_e\rangle_{CN}^{\infty}
-{[\langle\delta N_e\delta E\rangle_{CN}^{\infty}]^2\over {
\langle\delta^2 E\rangle_{CN}^{\infty}}}, 
\label{fin}
\eeq
where the superscript $\infty$ denotes $N\rightarrow\infty$. 
This worked beautifully for harmonic traps in various dimensions. 
These calculations, for our system, are also easily done for $\beta >1$. 
We now consider 
the spectrum (\ref{primes}) to be the infinite sequence of the primes, and 
evaluate the RHS of Eq.~(\ref{nav}). We readily obtain the convergent 
expression (for $\beta >1$)
\beq
\langle\delta^2 N_e\rangle_{MC}^{\infty}=
                        \sum_p {p^{\beta}\over {(p^{\beta}-1)^2}}-
{\left[\sum_p {(\ln p) p^{\beta}\over {(p^{\beta}-1)^2}}\right]^2\over 
{\sum_p {(\ln p)^2 p^{\beta}\over {(p^{\beta}-1)^2}}}}~.
\label{nav}
\eeq
The RHS of Eq.~(\ref{nav}) is nonzero, and therefore 
does not agree with the microcanonical result.
The failure of the above formalism of Navez {\it et al.}~\cite{navez} is 
because of the very special nature of the single-particle spectrum 
(\ref{primes}). Consider constructing the N-particle 
canonical partition function $Z_N(\beta)$ from this spectrum, as we had done. 
As $N\rightarrow \infty$, a little thought will show that 
$Z_N\rightarrow \zeta(\beta)$, where $\zeta(\beta)=\sum_n {1\over {n^\beta}}$ 
is the Riemann zeta function. This is because we are allowed to span over all 
$E$ in calculating $Z_N(\beta)$. Similarly, the grand partition function 
$\Xi_e(\alpha,\beta)$ with $\alpha=0$ is none other than the 
Euler product representation of the Riemann zeta function~\cite{edwards}. 
This has 
a density of states growing exponentially with $E$, and has been studied 
in connection with the limiting hadronic temperature~\cite{julia}. 
By contrast, the number of accessible states with a fixed $N$ and 
$E$ in the isolated microcanonical set-up does not increase at all. 
Since the energy remains locked in one microstate, the isolated system 
cannot be described through the usual concepts of statistical mechanics.  

This research is supported by NERSC (Canada). The authors would like to 
thank David M.~Cooke and Jamal Sakhr for helpful discussions.

\newpage
\begin{figure} 
\caption{a) Average occupancy in the ground state $N_0/N$ versus temperature $T$ for $N=100$ in the canonical and grand canonical ensembles. b) Plot of the relative ground state number fluctuation in both ensembles.  Note the steep rise in the grand canonical fluctuation.}
\label{figure1}
\vspace{7 mm}
\caption{a) Plot of the average occupancy in the excited states $N_{ex}/N$, for $N=100$, versus the excitation energy $E_x$ in the canonical ensemble (continuous bold curve), compared with the exact microcanonical calculation. The latter calculation is done for $E_x=\ln n$, where $n$ is an integer, and the results are shown by dark points. These are joined by dotted lines to emphasize their zigzag character. For example, the sixth point (including $0$) corresponds to $E_x=\ln 6$, and gives $N_e=2$, corresponding to the prime factor decomposition $2\times 3$. }
\vspace{7 mm}
\caption{Plot of the canonical entropy as a function of temperature $T$ on the left, and excitation energy $E_x$ on the right, for $N=100$. The microcanonical entropy is zero. }
\end{figure}

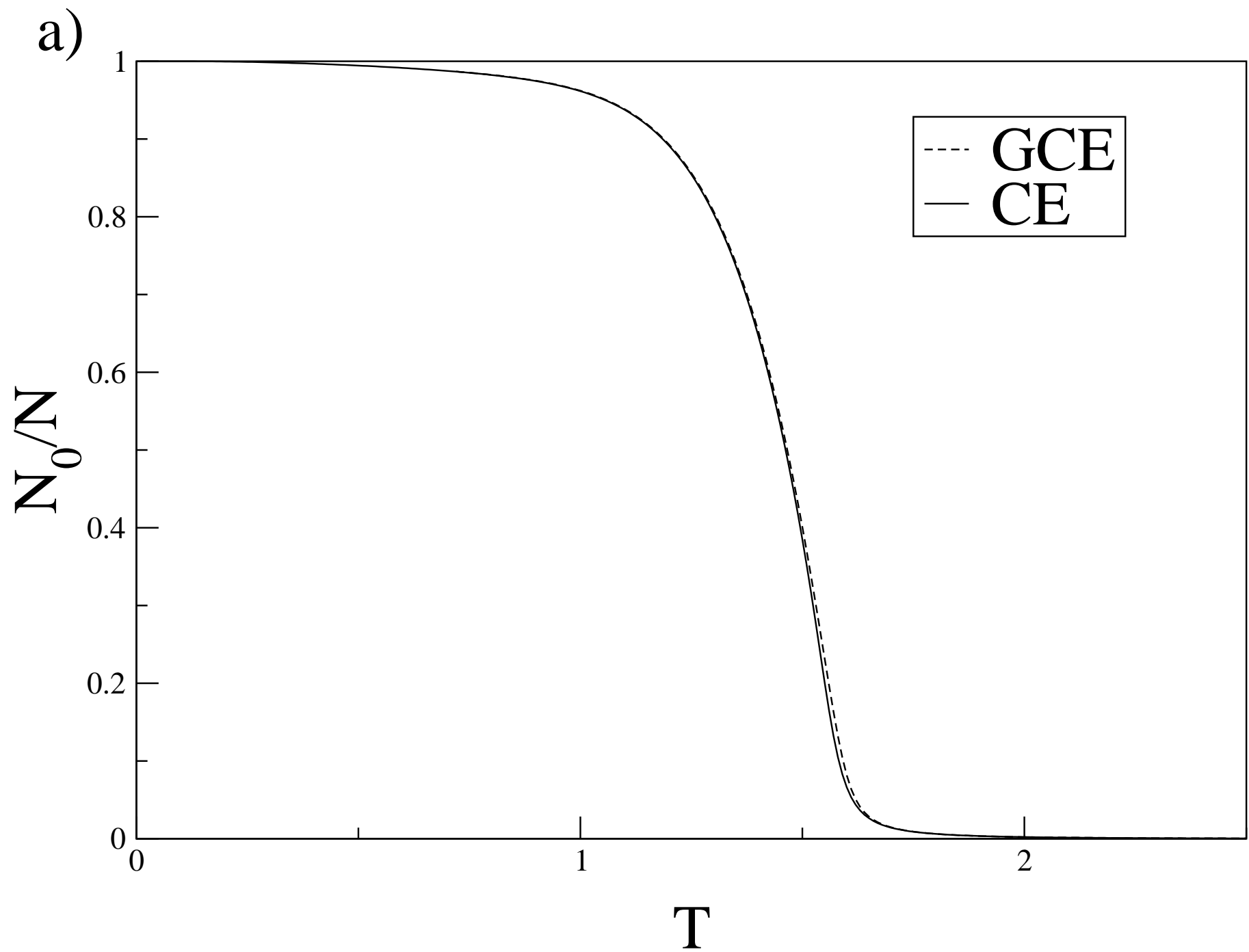
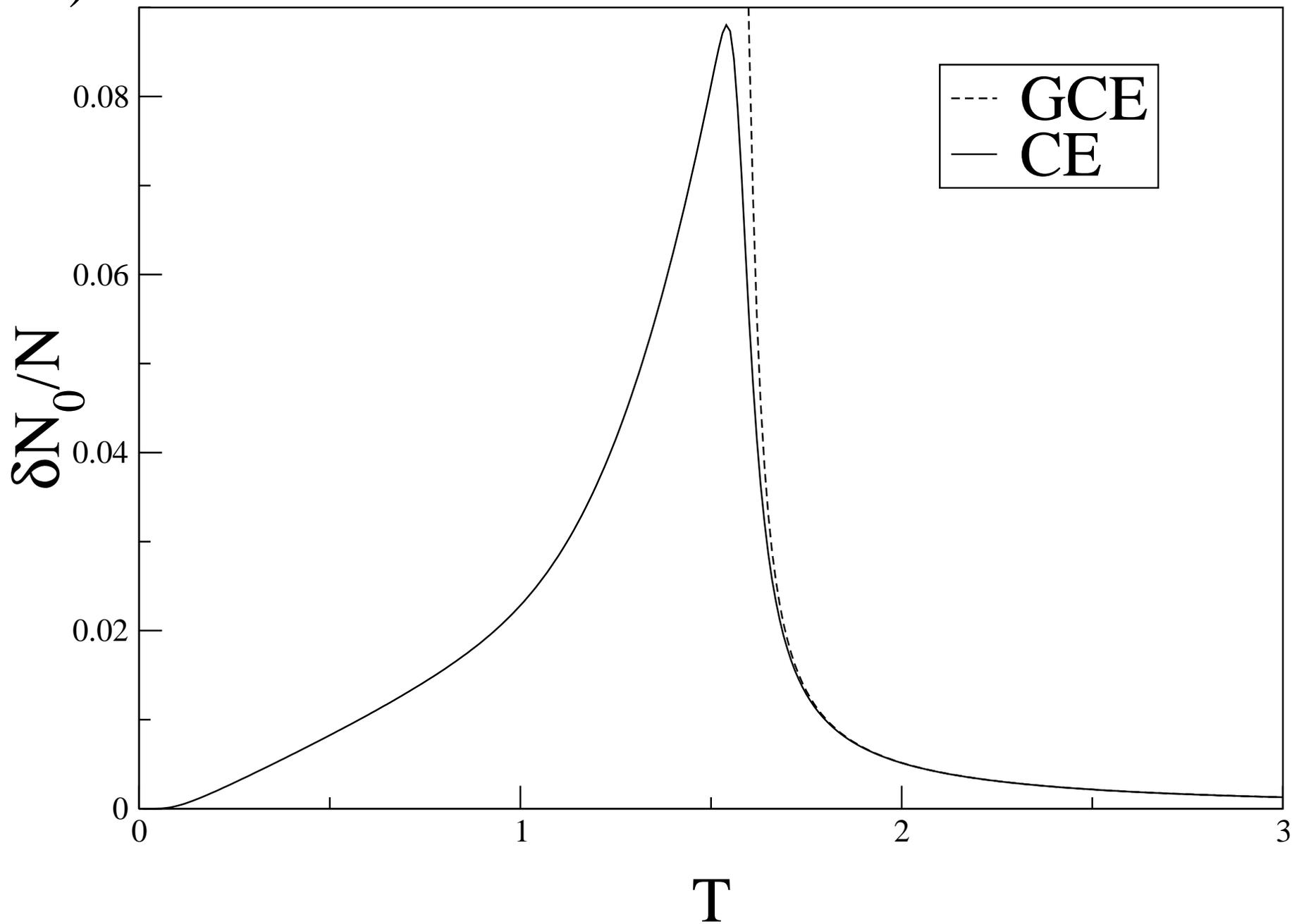
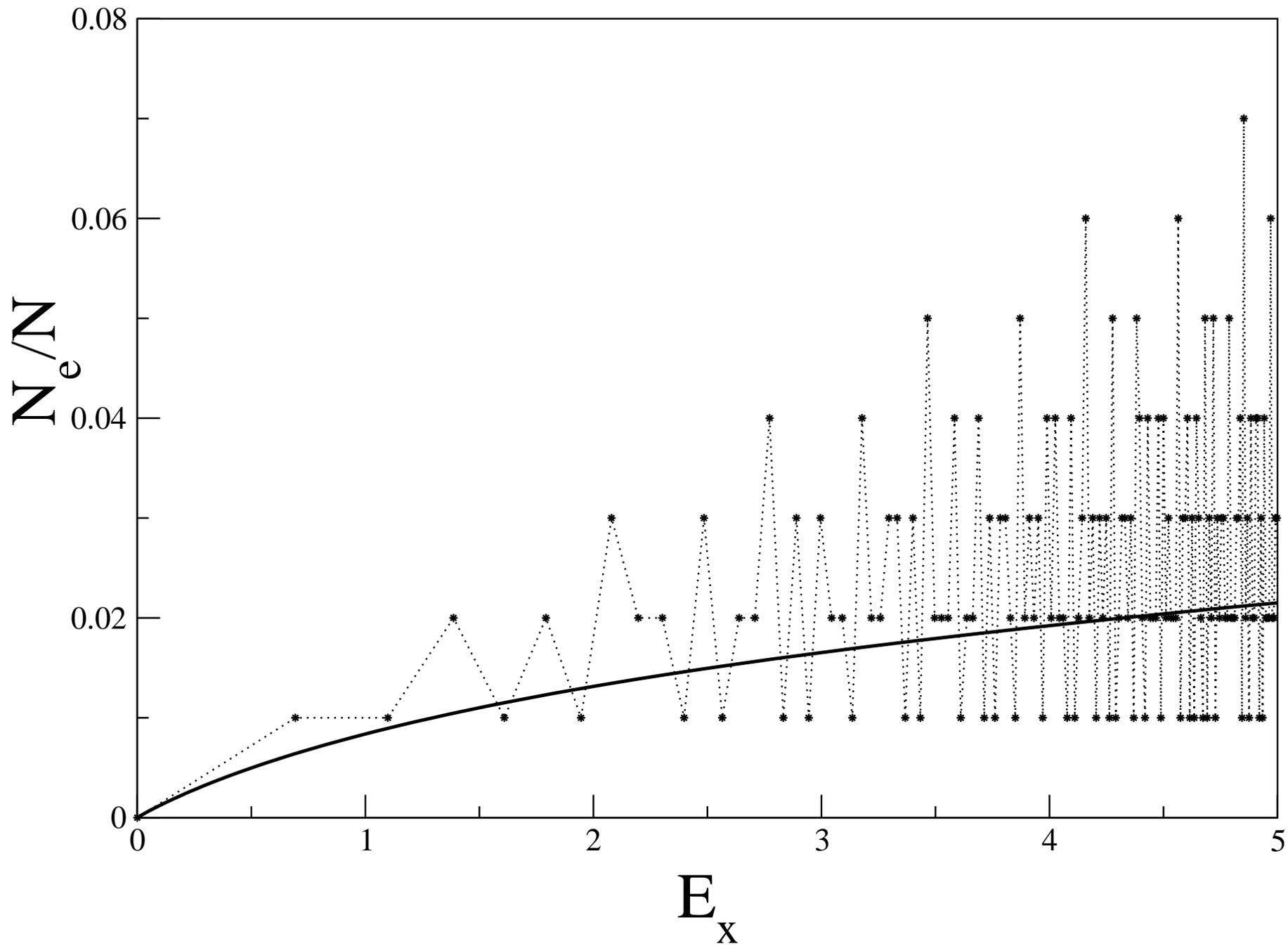
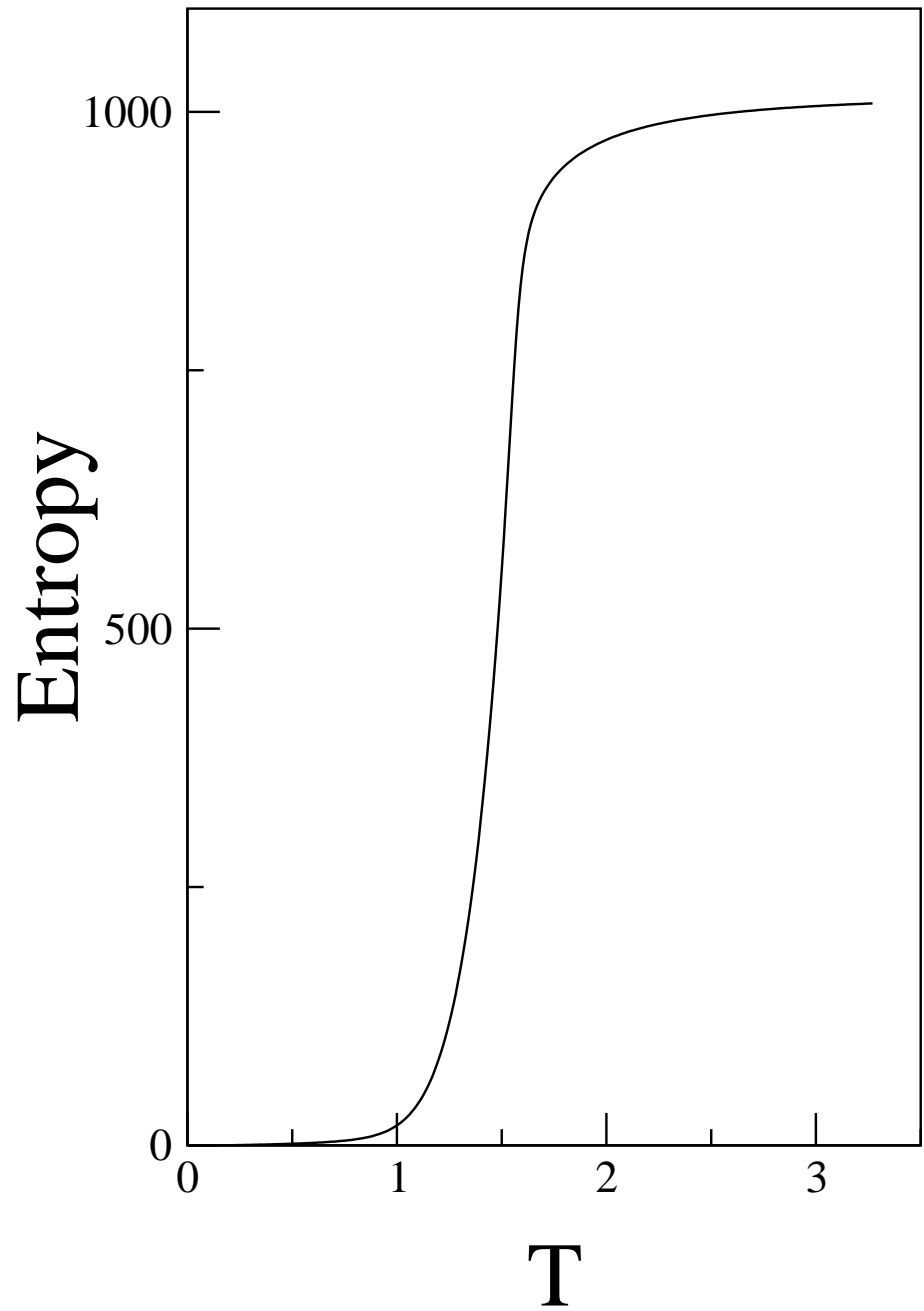
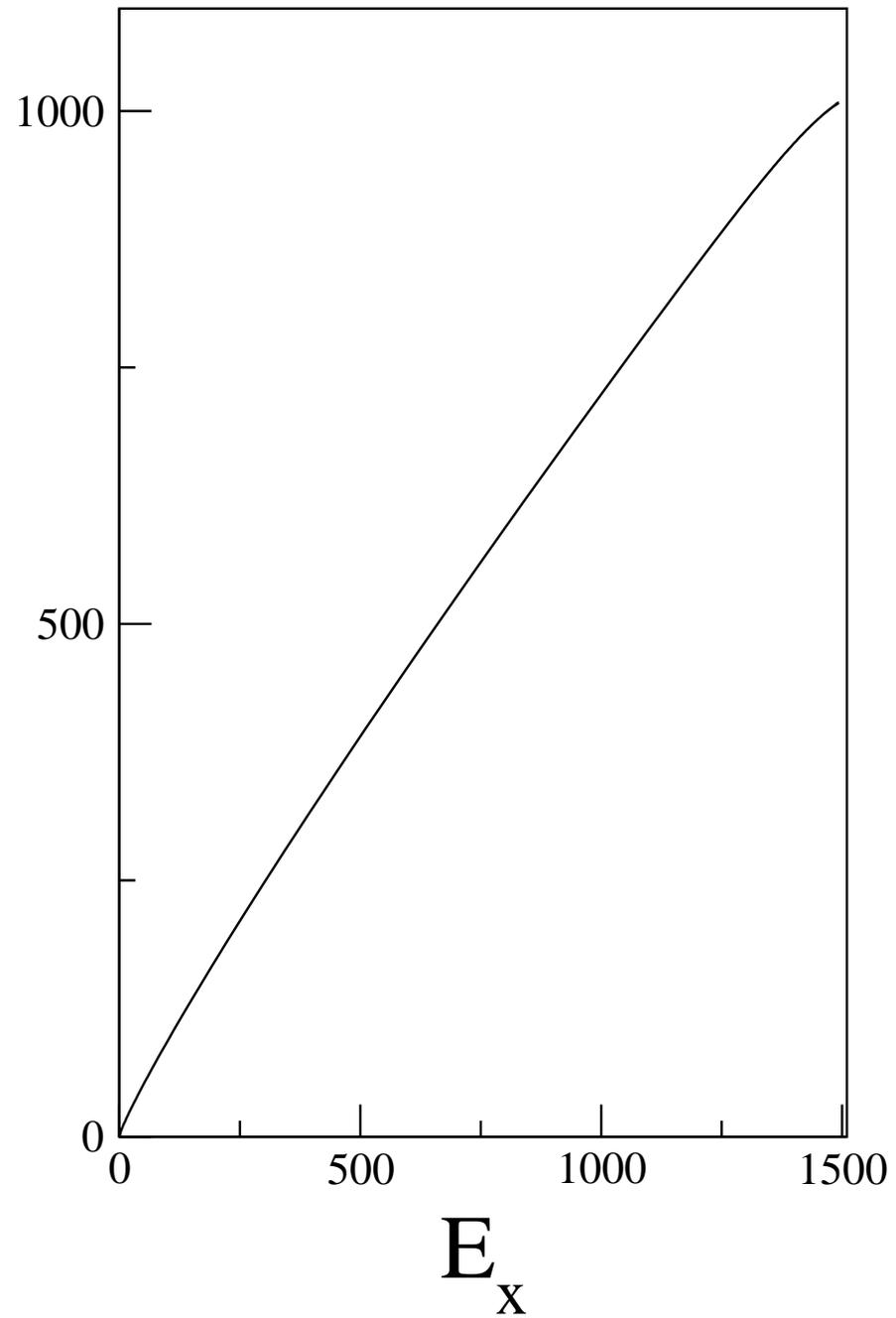
\end{document}